\documentclass[journal]{IEEEtran}
\usepackage{cite}
\ifCLASSINFOpdf
\else
  \usepackage[dvips]{graphicx}
  \usepackage{subfigure}
\fi
\begin{document}
\title{P-Type Tunnel FETs With Triple Heterojunctions}

\author{Jun~Z.~Huang,
        Pengyu~Long,
        Michael~Povolotskyi,
        Gerhard~Klimeck,
        and~Mark~J.~W.~Rodwell% <-this % stops a space
\thanks{J. Z. Huang, P. Long, M. Povolotskyi, and G. Klimeck are with
the Network for Computational Nanotechnology and Birck Nanotechnology
Center, Purdue University, West Lafayette, IN 47907 USA (e-mail:
junhuang1021@gmail.com).}% <-this % stops a space
\thanks{M. J. W. Rodwell is with the Department of Electrical and Computer
Engineering, University of California at Santa Barbara, Santa Barbara,
CA 93106-9560 USA.}% <-this % stops a space
\thanks{Manuscript received xxx xx, 2016; revised xxxx xx, 2016. This work uses nanoHUB.org
computational resources operated by the Network for Computational
Nanotechnology funded by the U.S. National Science Foundation
under Grant EEC-0228390, Grant EEC-1227110, Grant EEC-0634750,
Grant OCI-0438246, Grant OCI-0832623, and Grant OCI-0721680. This
material is based upon work supported by the National Science Foundation
under Grant 1125017. NEMO5 developments were critically supported by an
NSF Peta-Apps award OCI-0749140 and by Intel Corp.}}

\maketitle

\begin{abstract}
A triple-heterojunction (3HJ) design is employed to improve p-type InAs/GaSb heterojunction (HJ) tunnel FETs.
The added two HJs (AlInAsSb/InAs in the source and GaSb/AlSb in the channel) significantly
shorten the tunnel distance and create two resonant states, greatly improving the ON state tunneling probability.
Moreover, the source Fermi degeneracy is reduced by the increased source (AlInAsSb) density of states and the OFF state leakage is reduced by the heavier channel (AlSb) hole effective masses. Quantum ballistic transport simulations show, that with $V_{\rm{DD}}=0.3\rm{V}$ and $I_{\rm{OFF}}=10^{-3}\rm{A/m}$, $I_{\rm{ON}}$ of $582\rm{A/m}$ ($488\rm{A/m}$) is obtained at $30\rm{nm}$ ($15\rm{nm}$) channel length, which is comparable to n-type 3HJ counterpart and significantly exceeding p-type silicon MOSFET. Simultaneously, the nonlinear turn on and delayed saturation in the output characteristics are also greatly improved.
\end{abstract}


% \fi
\IEEEpeerreviewmaketitle

\section{Introduction}
\IEEEPARstart{S}{teep} subthreshold swing (SS) devices, such as tunnel field-effect transistors (TFETs), offer great potential in building future low-power integrated circuits. One problem of TFETs is the low tunneling probability hence low ON state current ($I_{\rm{ON}}$). To achieve large $I_{\rm{ON}}$, III-V TFET designs have been intensively studied \cite{ionescu2011tunnel}. In particular, InAs/GaSb HJ TFETs can considerably boost $I_{\rm{ON}}$ due to their broken/staggered band alignments \cite{mohata2011demonstration}. However, under strong confinement, required for good electrostatic control, the effective band gap and transport effective masses both increase, seriously limiting the tunneling probability. Methods to improve InAs/GaSb HJ n-type TFETs (nTFETs) include strain and doping engineering \cite{brocard2013design,Verreck2016}, resonant enhancement \cite{avci2013heterojunction,pala2015exploiting,Long2016design}, and source/channel heterojunctions \cite{Ganapathi2011,brocard2014large,Li2015,Long2016,Long2016drc}. For p-type TFETs (pTFETs), the problem is more severe, as the optimal source doping density is limited by the small conduction band density of states (DOS) \cite{Knoch2010}. This leads to a large depletion region in the source and thus, smaller $I_{\rm{ON}}$ than nTFETs \cite{Avci2011,Avci2015,Huang2015}. Doping and heterojunction engineering in the source \cite{Verreck2014} have been proposed to mitigate this problem. Another problem of TFETs is the superlinear onset and delayed saturation of the output characteristics. It has been shown that a large channel DOS degrades the output characteristics through large channel inversion charge \cite{Taur2015,Wu2016}. This is particularly relevant for pTFETs since the valence band DOS of most III-V materials is very large. These two issues make it very challenging to build complementary III-V TFET logic, which requires both high-performance nTFETs and pTFETs. Wu {\it et al.} \cite{Wu2016} note that the required source and channel materials for HJ nTFETs and pTFETs differ greatly.

For HJ nTFETs, it has been previously shown that better ON/OFF ratio is achieved by adopting (1$\bar{1}$0)/[110] as the confinement/transport crystal orientation, because smaller tunnel barrier energy and transport effective masses are found in this orientation \cite{Long2016}. It has been further shown that the ballistic $I_{\rm{ON}}$ can be greatly increased by adding two more HJs, one in the channel \cite{Long2016} and one in the source, so as to form a 3HJ design \cite{Long2016drc}. In this paper, we show that by crystal orientation engineering, using the 3HJ design, we can also solve the above mentioned problems of pTFETs, achieving very large ballistic $I_{\rm{ON}}$ as well as improved output I-V characteristics.

\section{Heterojunction (HJ) pTFET}
\begin{figure}[htbp] \centering
{\includegraphics[width=4.35cm]{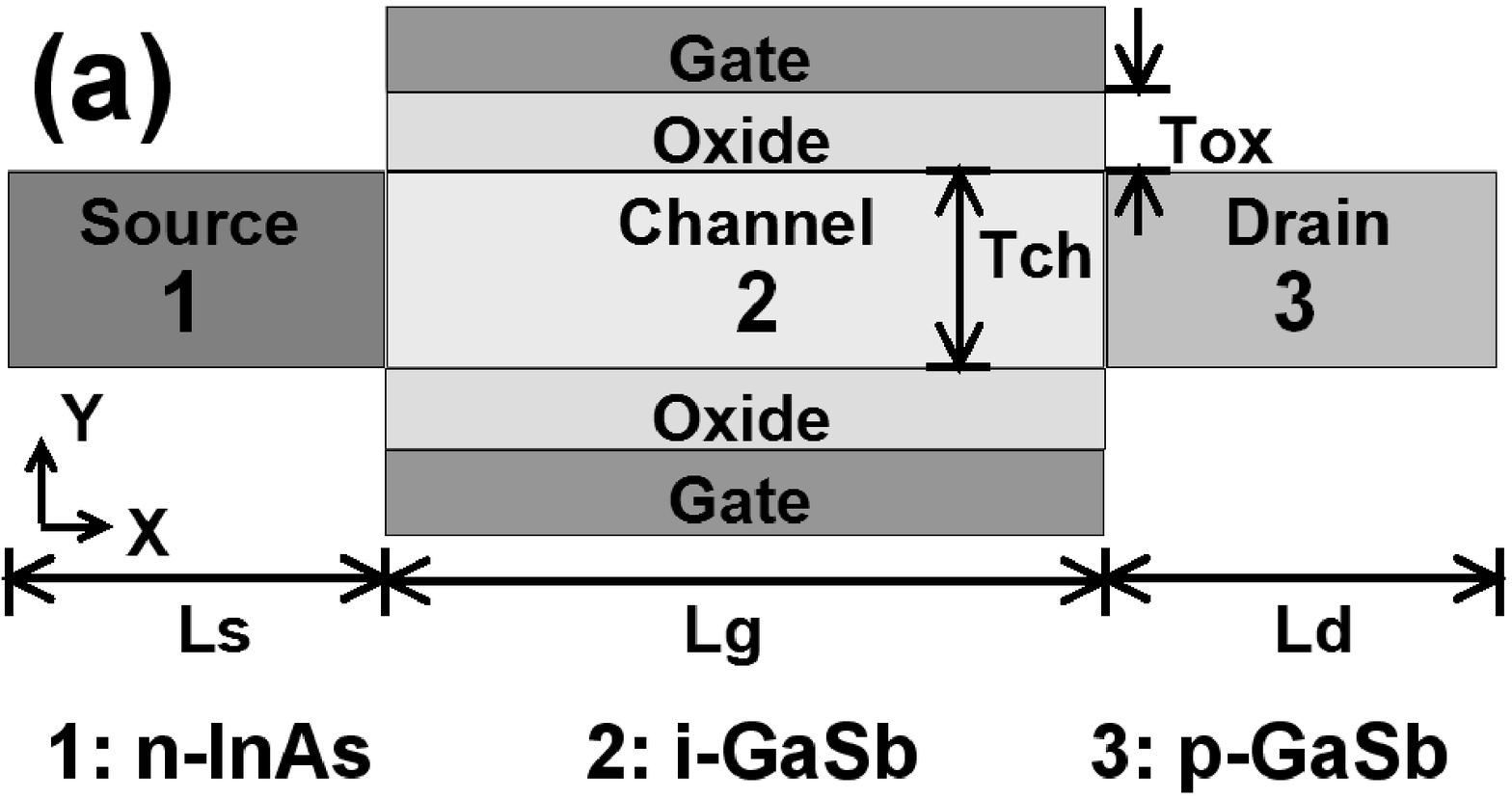}}
{\includegraphics[width=4.35cm]{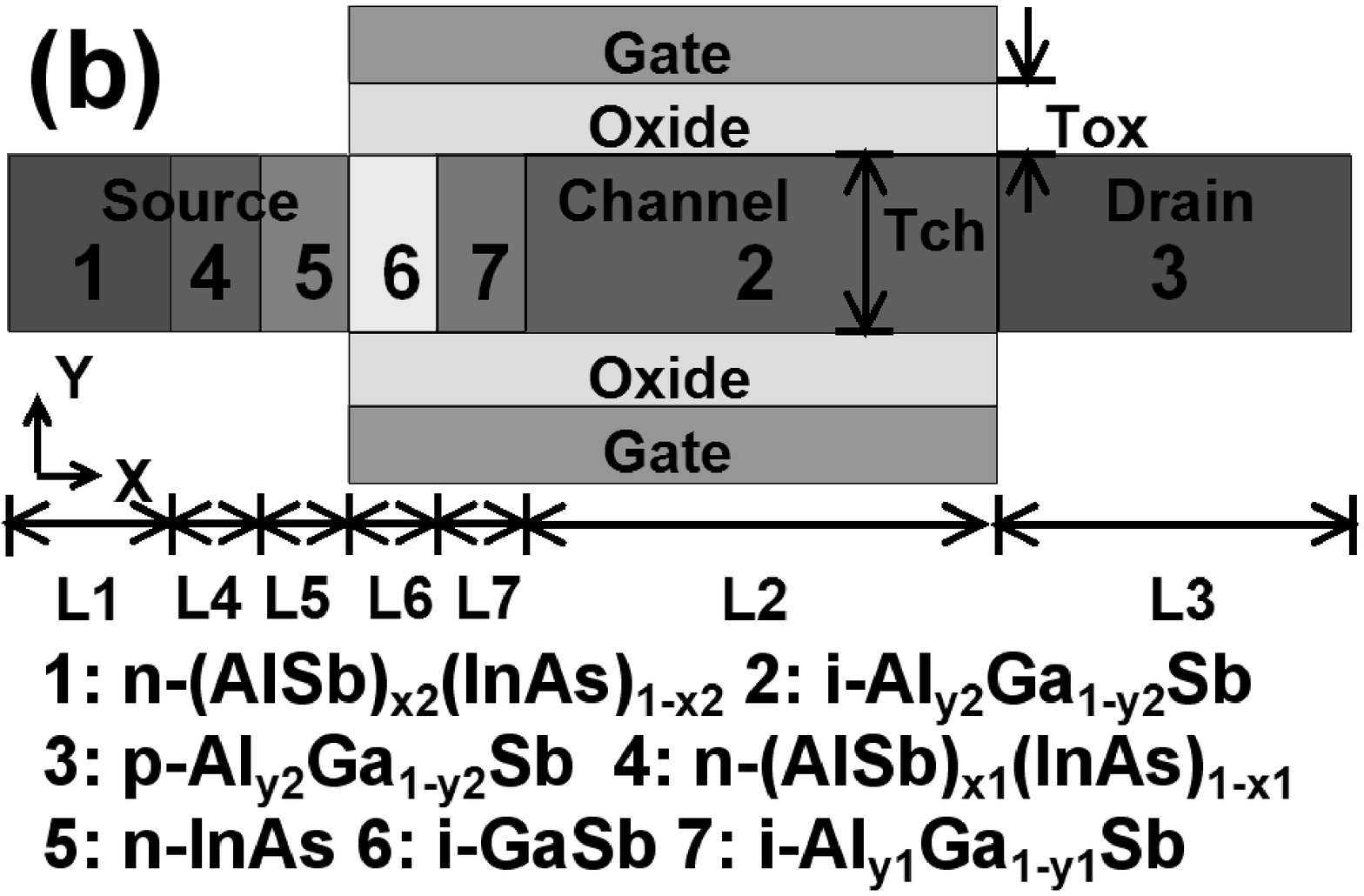}}
\caption{Device structures and material compositions of a HJ pTFET (a) and a 3HJ pTFET (b).}
\label{fig:device}
\end{figure}

The ultra-thin-body (UTB) HJ pTFET consists of an InAs source and a GaSb channel/drain (Fig. \ref{fig:device} (a)), with the device parameters listed in Table \ref{tab:device_param}. The NEMO5 tool \cite {Steiger2011} is used to simulate the device by solving Poisson equation and open boundary Schr\"{o}dinger equation \cite{Luisier2006} self-consistently. The device Hamiltonian is described by transferrable full-band tight binding (TB) scheme ($sp^3d^5s^*$ basis including spin-orbit coupling) \cite{Tan2015}, whose parameters at 300K are taken from \cite{Tan2016}.

\begin{table}
\caption{Device parameters. $Dx$ denotes the doping density of region $x$.}
\label{tab:device_param}       % Give a unique label
\begin{tabular}{p{1.5cm}p{1.5cm}p{0.8cm}p{0.8cm}p{0.8cm}p{0.8cm}}
\hline\noalign{\smallskip}
Ls &Lg &Ld & Tch & Tox & $\epsilon_{ox}$\\
15nm  & 15nm  & 10nm & 1.8nm &1.8nm  & 9.0\\
\hline\noalign{\smallskip}
$D1$ ($\rm{cm}^{-3}$) & $D4$ ($\rm{cm}^{-3}$) &L4 &L5 &L6 & L7 \\
$+1\times 10^{19}$ & $+5\times 10^{19}$  & 2.0nm  & 3.3nm  & 1.7nm & 2.0nm \\
\hline\noalign{\smallskip}
$D5$ ($\rm{cm}^{-3}$) &$D3$ ($\rm{cm}^{-3}$) &x2 & x1 & y1 & y2\\
$+5\times 10^{19}$  &$-5\times 10^{19}$  & 0.23& 0.12 & 0.5 & 1.0\\
\hline\noalign{\smallskip}
\end{tabular}
\end{table}

\begin{figure}[htbp] \centering
{\includegraphics[width=4.35cm]{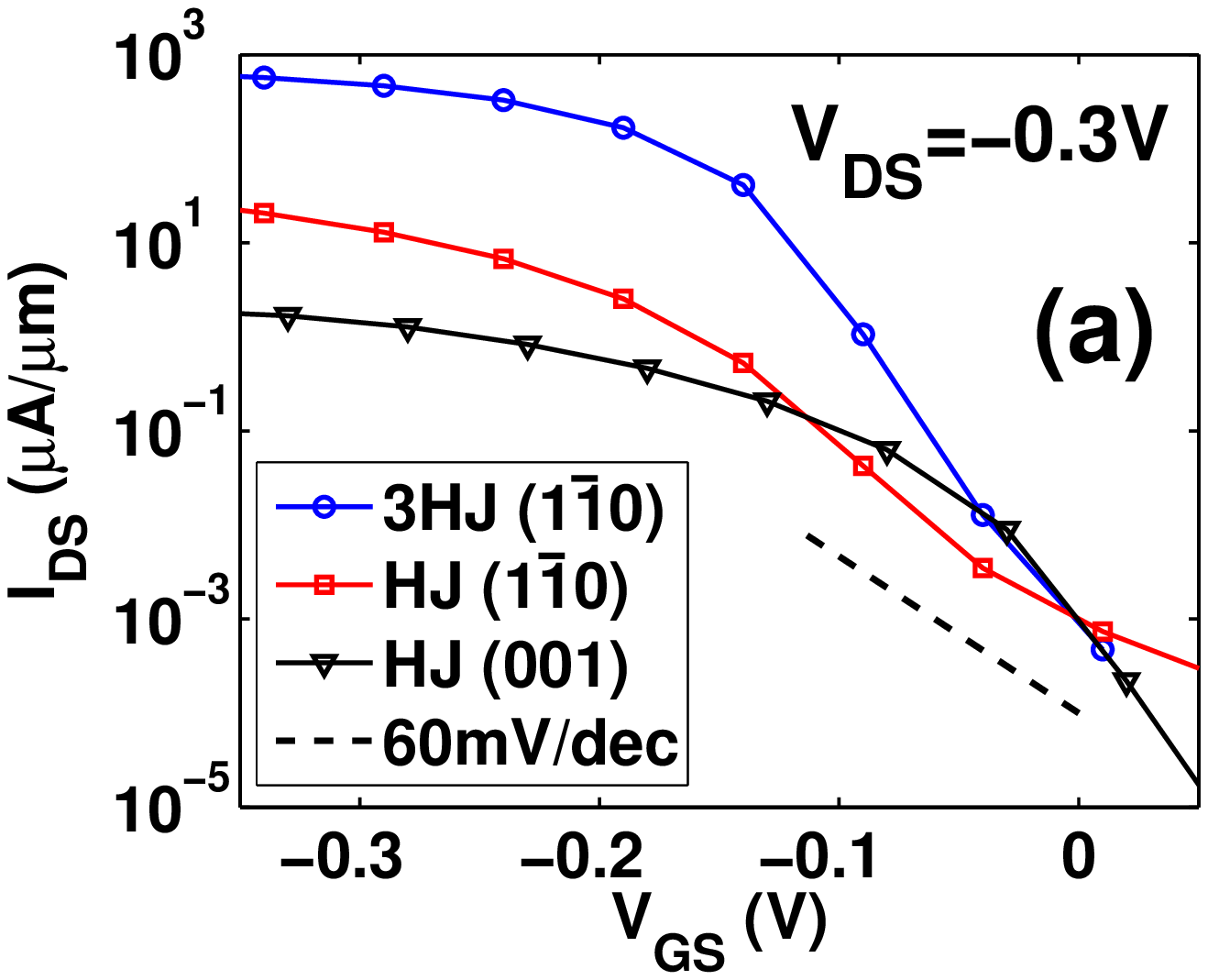}}
{\includegraphics[width=4.35cm]{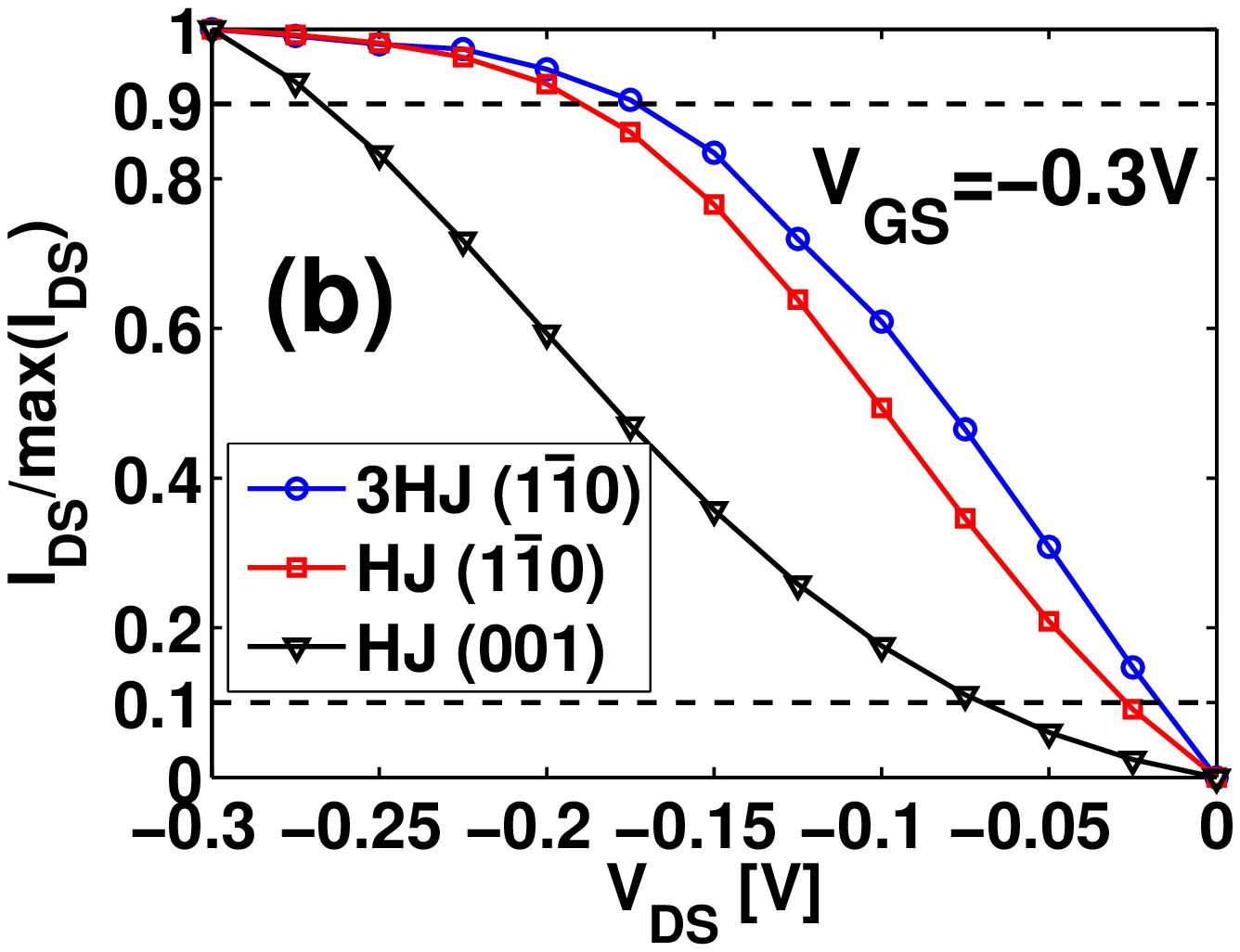}}
{\includegraphics[width=4.35cm]{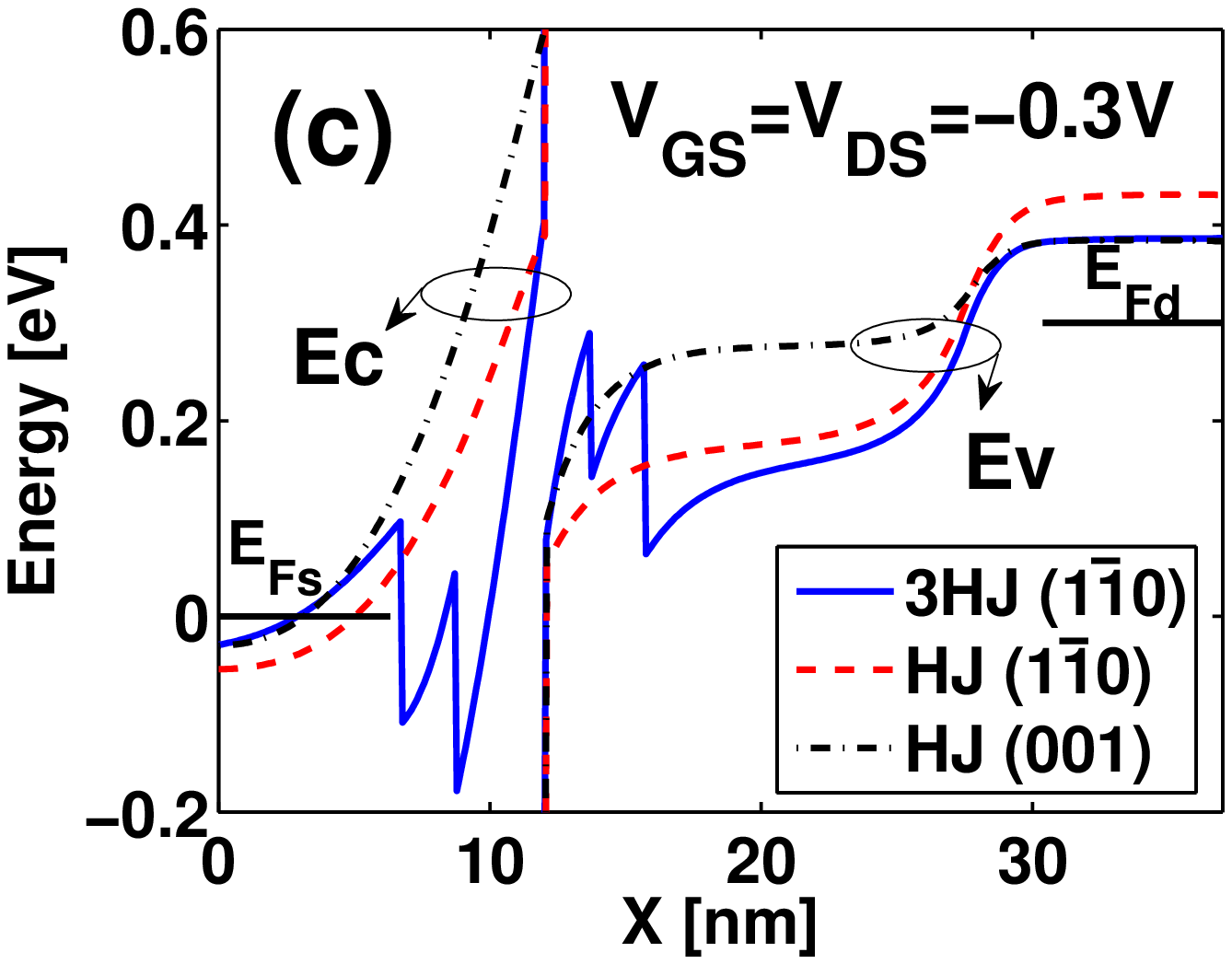}}
{\includegraphics[width=4.35cm]{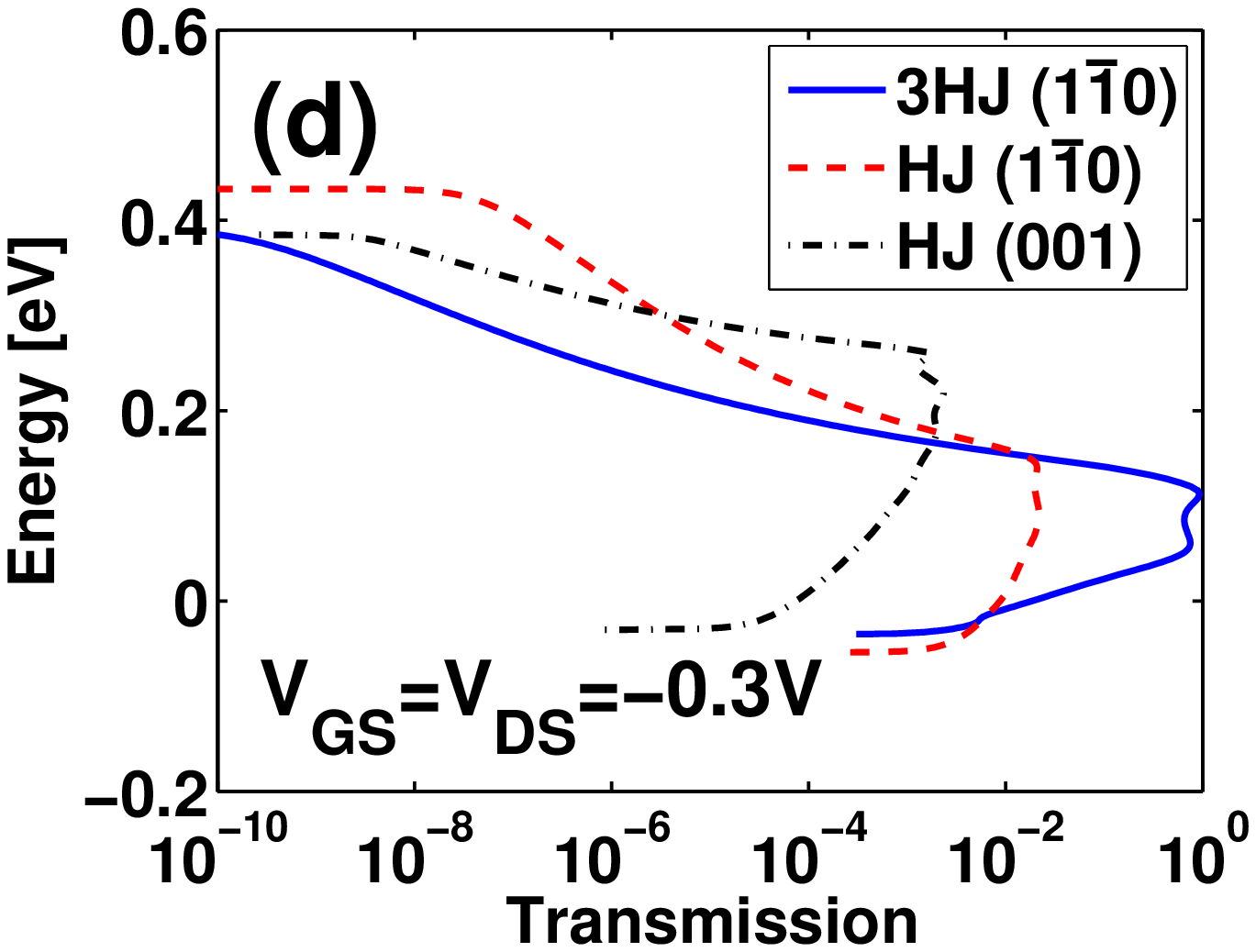}}
\caption{(a) Transfer characteristics, (b) output (normalized) characteristics, (c) band diagrams, and (d) transmission probabilities of two HJ pTFETs, in the (001)/[100] and (1$\bar{1}$0)/[110] orientation, respectively, and one 3HJ pTFET in the (1$\bar{1}$0)/[110] orientation. Ec (Ev): conduction (valence) band edge. $\rm{E}_{\rm{Fs}}$ ($\rm{E}_{\rm{Fd}}$): source (drain) Fermi level.}
\label{fig:iv_band_trans}
\end{figure}

The (1$\bar{1}$0)/[110] orientation performs better than the (001)/[100] orientation. As compared in Fig. \ref{fig:iv_band_trans} (a), with $V_{\rm{DD}}=0.3\rm{V}$ and $I_{\rm{OFF}}=10^{-3}\rm{\mu A/\mu m}$, $I_{\rm{ON}}$ is 14.5$\rm{\mu A/\mu m}$ in the (1$\bar{1}$0)/[110] orientation. While in the (001)/[100] orientation $I_{\rm{ON}}$ is only 1.4$\rm{\mu A/\mu m}$ although the SS is better. The (1$\bar{1}$0)/[110] orientation not only improves $I_{\rm{ON}}$ but also improves the superlinear onset and delayed saturation of the $I_{DS}$-$V_{DS}$ characteristics. As compared in Fig. \ref{fig:iv_band_trans} (b), the onset and saturation voltages, defined here as the drain voltages corresponding to 10\% and 90\% of the maximum drain current, are both reduced in the (1$\bar{1}$0)/[110] orientation.

\begin{figure}[htbp] \centering
{\includegraphics[width=4.35cm]{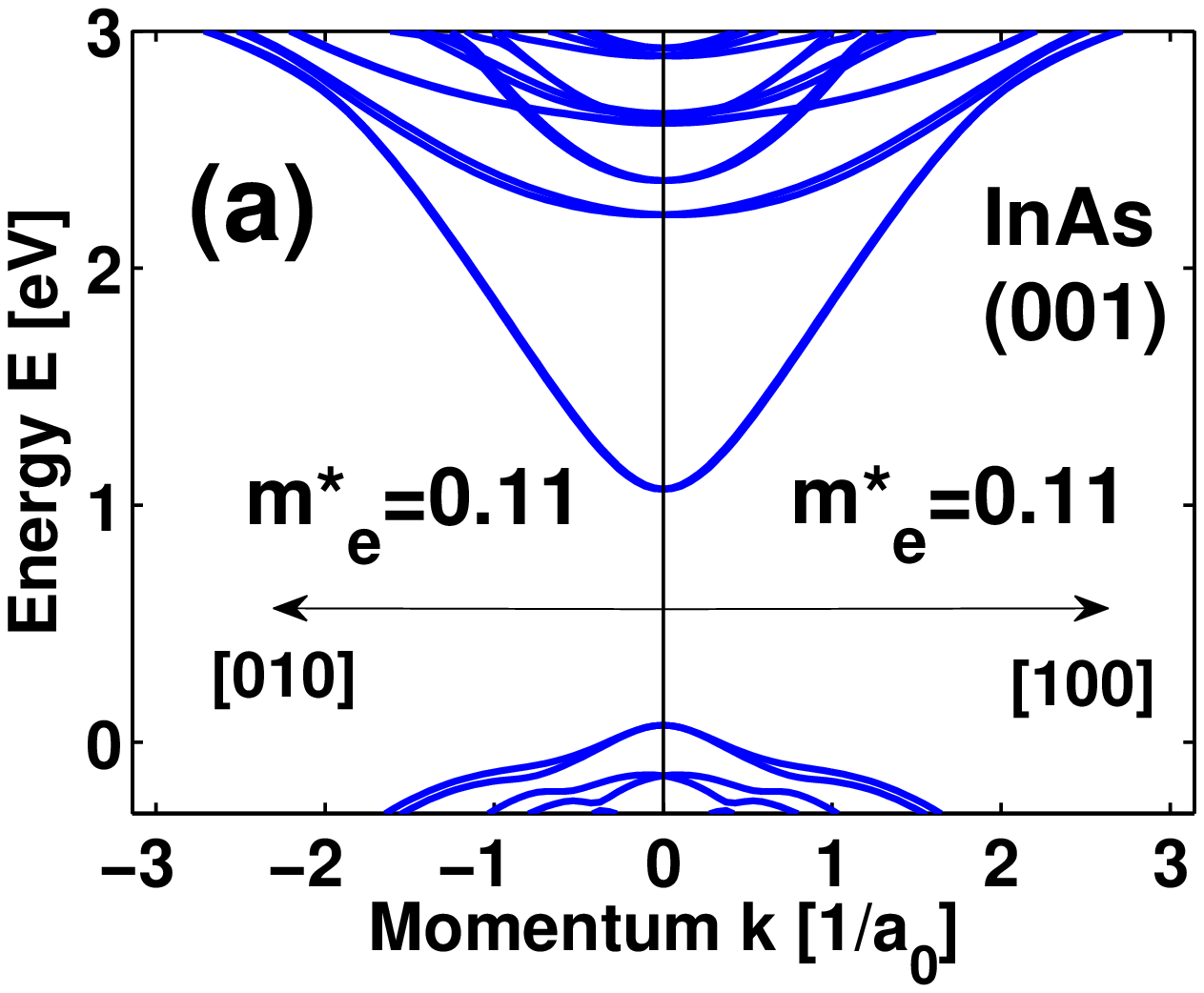}}
{\includegraphics[width=4.35cm]{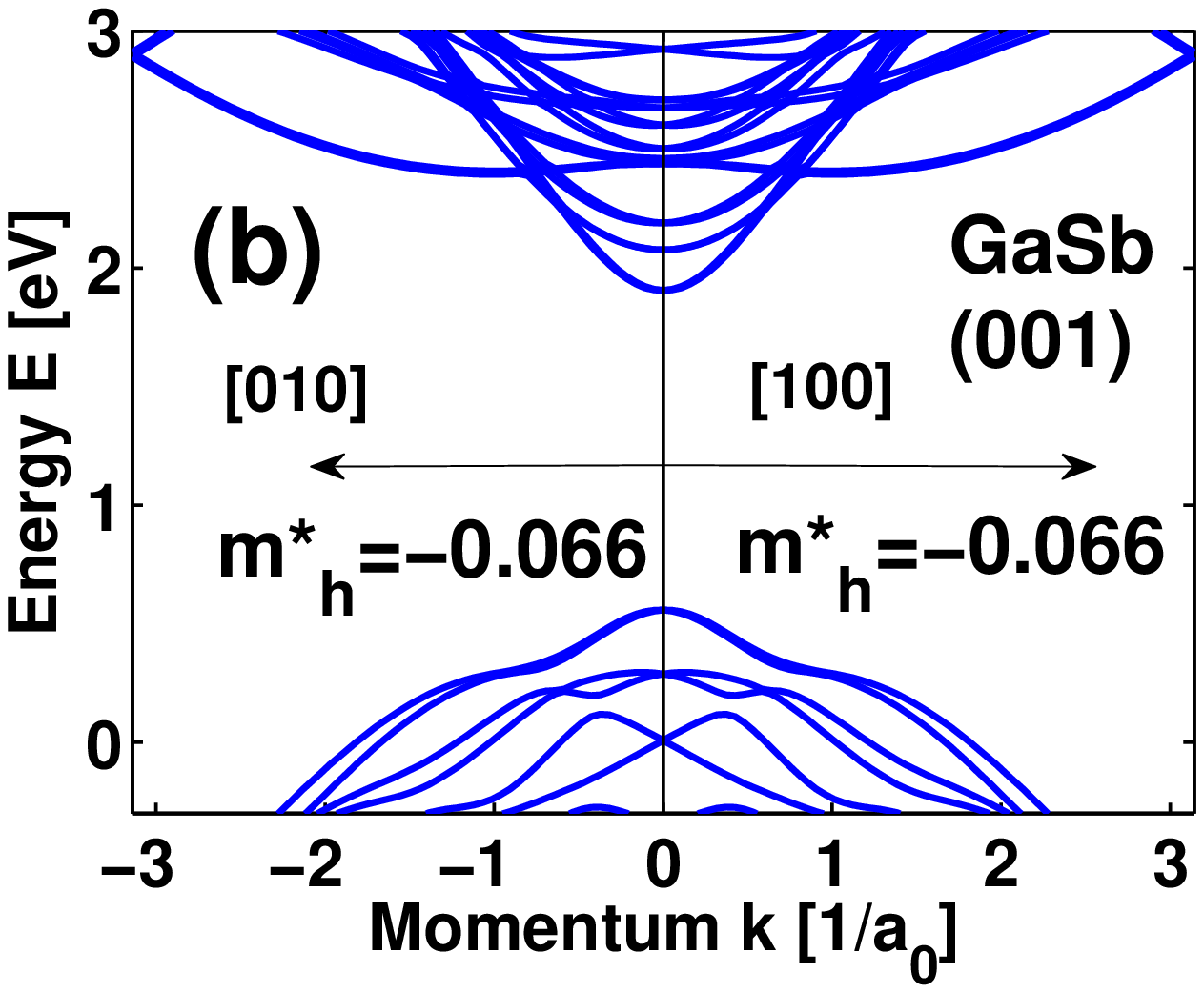}}
{\includegraphics[width=4.35cm]{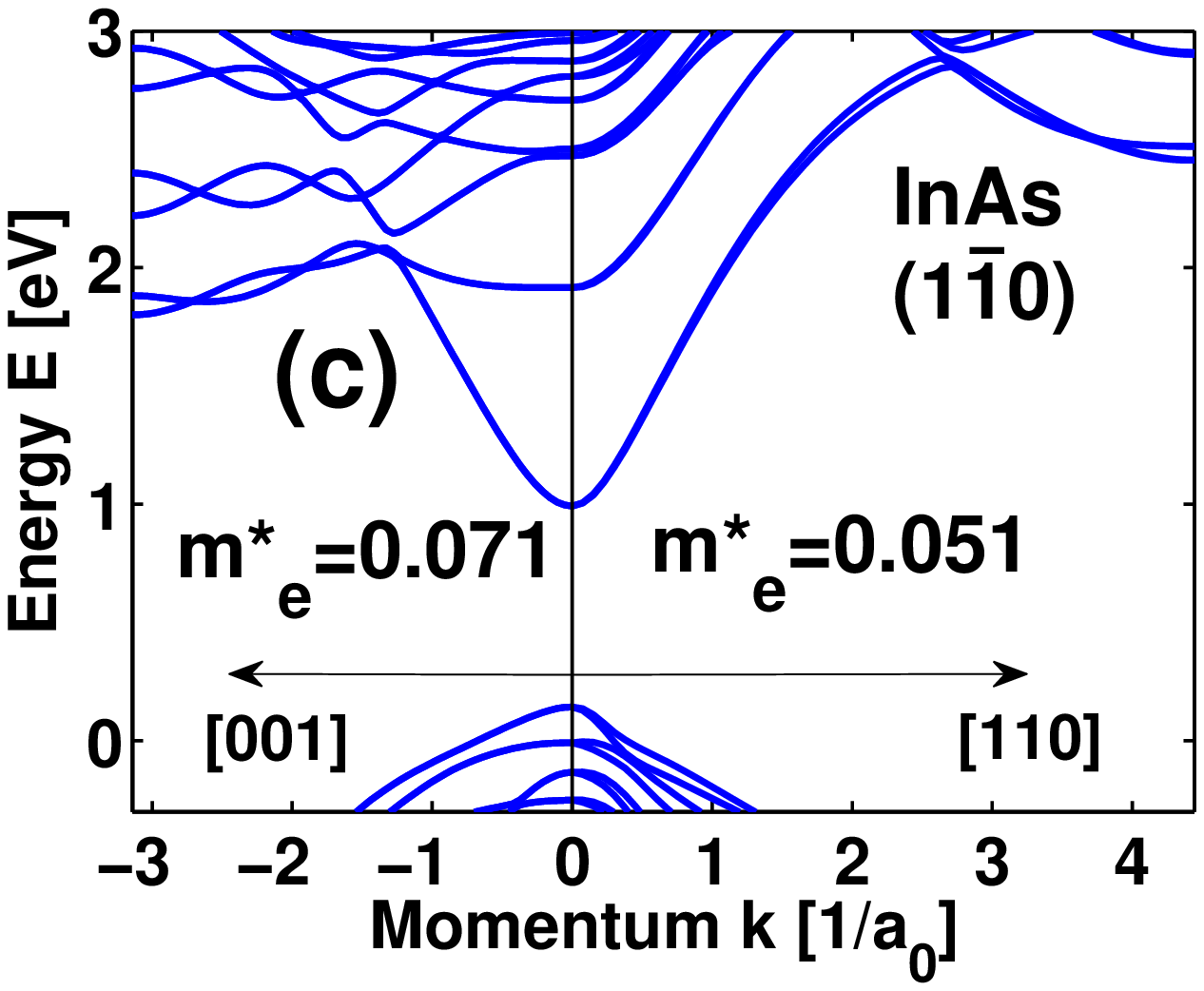}}
{\includegraphics[width=4.35cm]{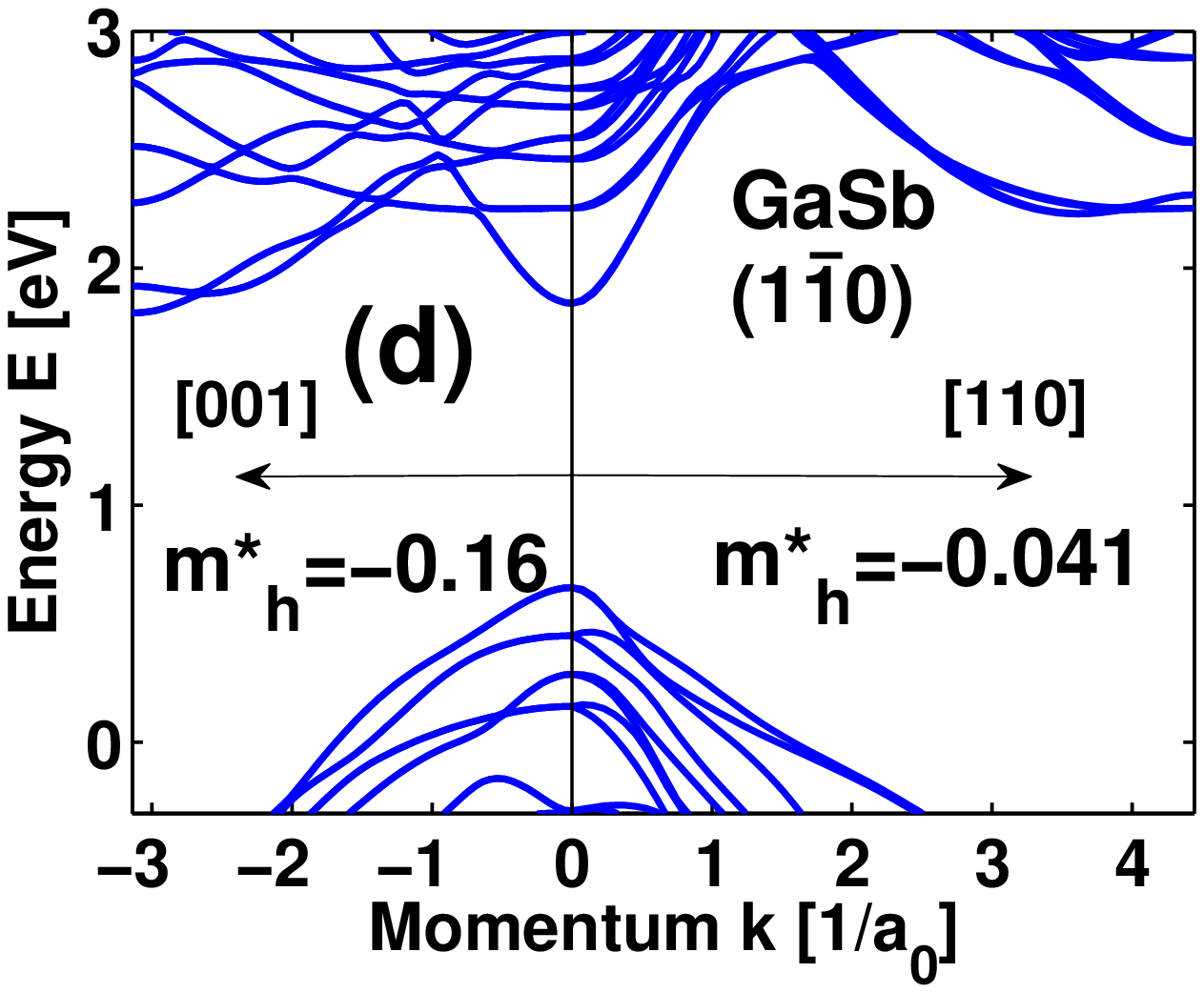}}
{\includegraphics[width=4.35cm]{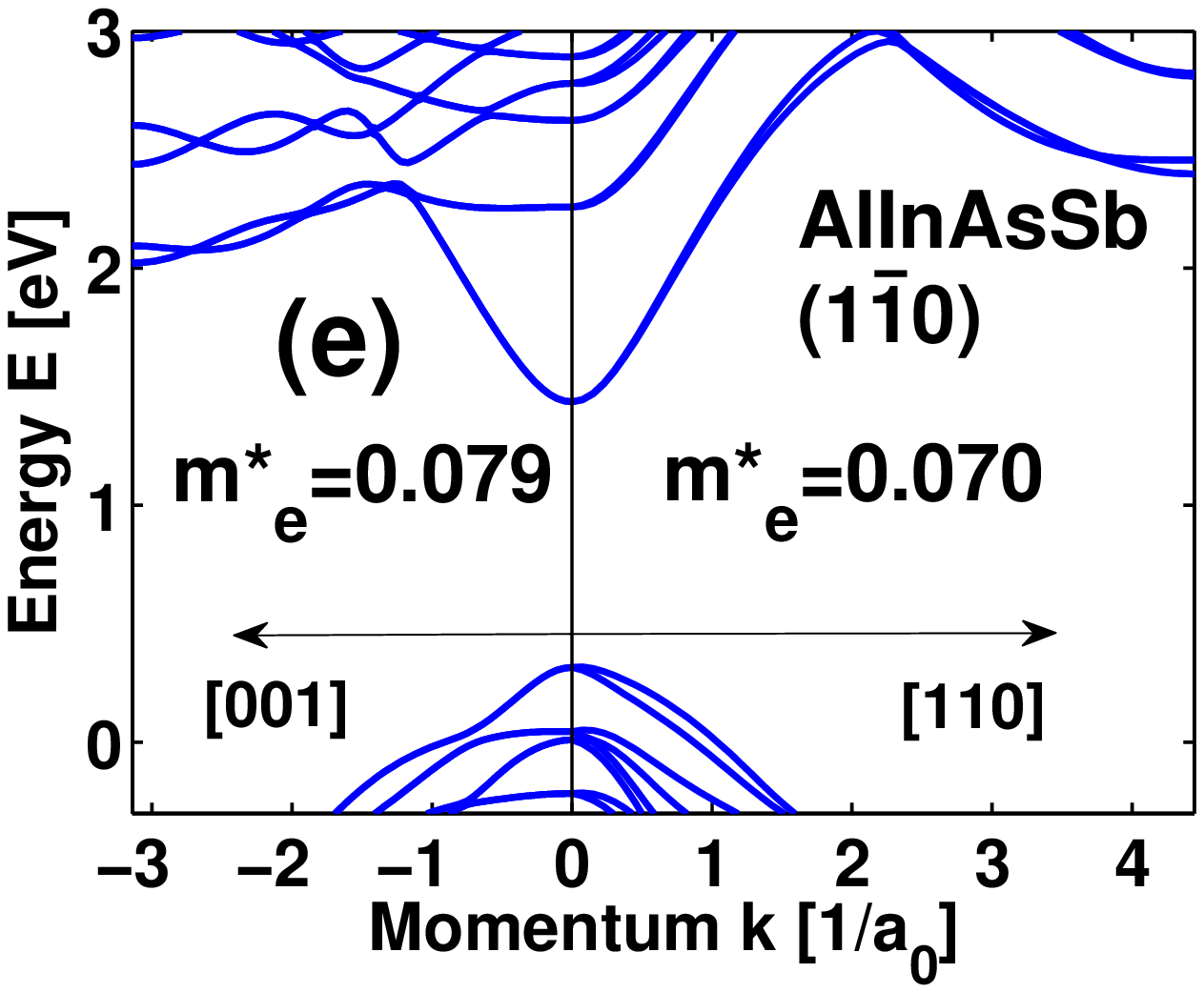}}
{\includegraphics[width=4.35cm]{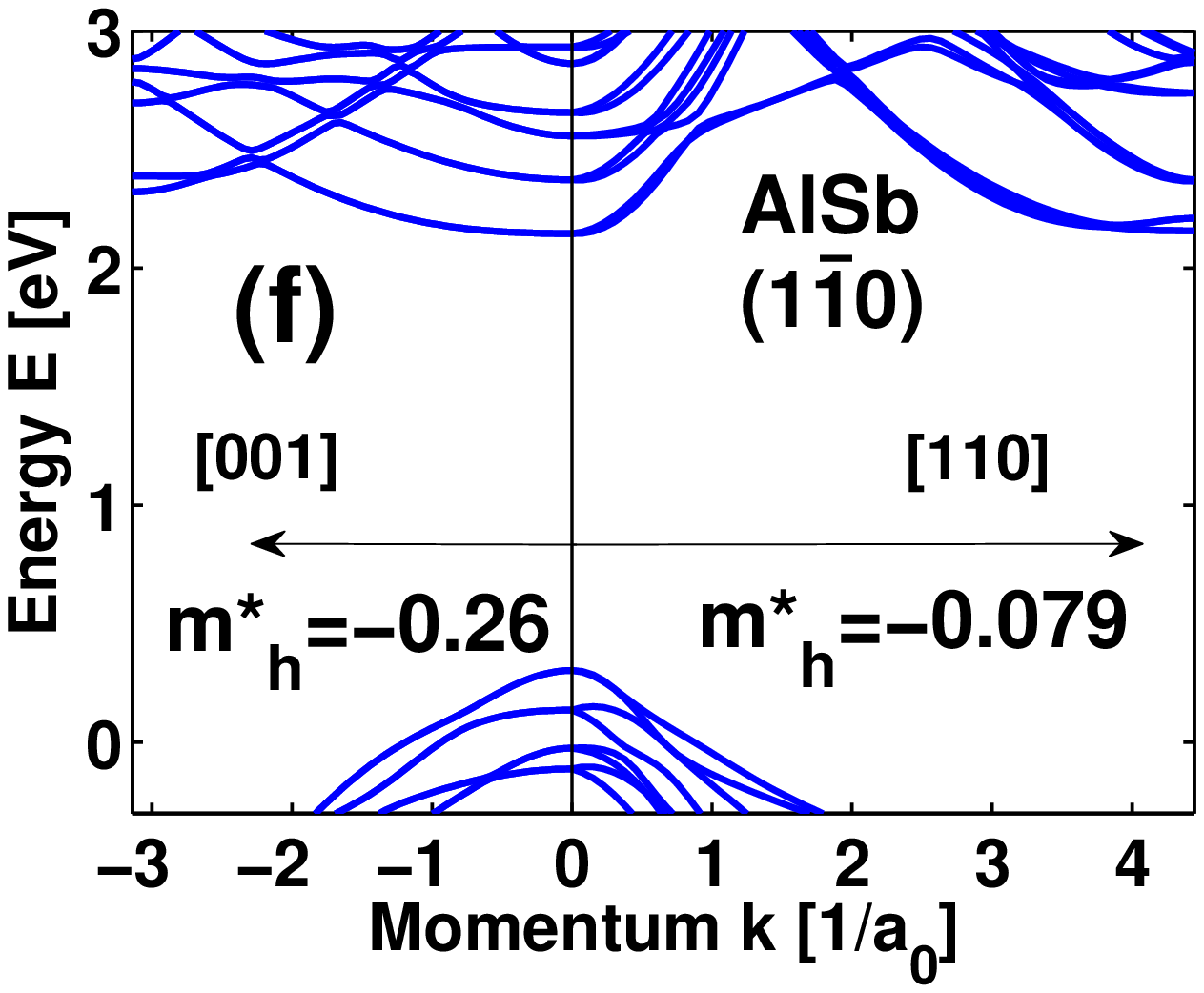}}
\caption{$E$-$k$ diagrams for 1.8nm thick UTBs with (001) InAs (a), (001) GaSb (b), (1$\bar{1}$0) InAs (c), (1$\bar{1}$0) GaSb (d), (1$\bar{1}$0) $\rm{(AlSb)}_{0.23}\rm{(InAs)}_{0.77}$ (e), and (1$\bar{1}$0) AlSb (f), in the transport direction (positive $k$) and transverse direction (negative $k$). The $a_0$ is the lattice constant. The $m^*_e$ ($m^*_h$) is the electron (hole) effective mass at the band edge.}
\label{fig:ek}
\end{figure}

\begin{figure}[htbp] \centering
{\includegraphics[width=4.35cm]{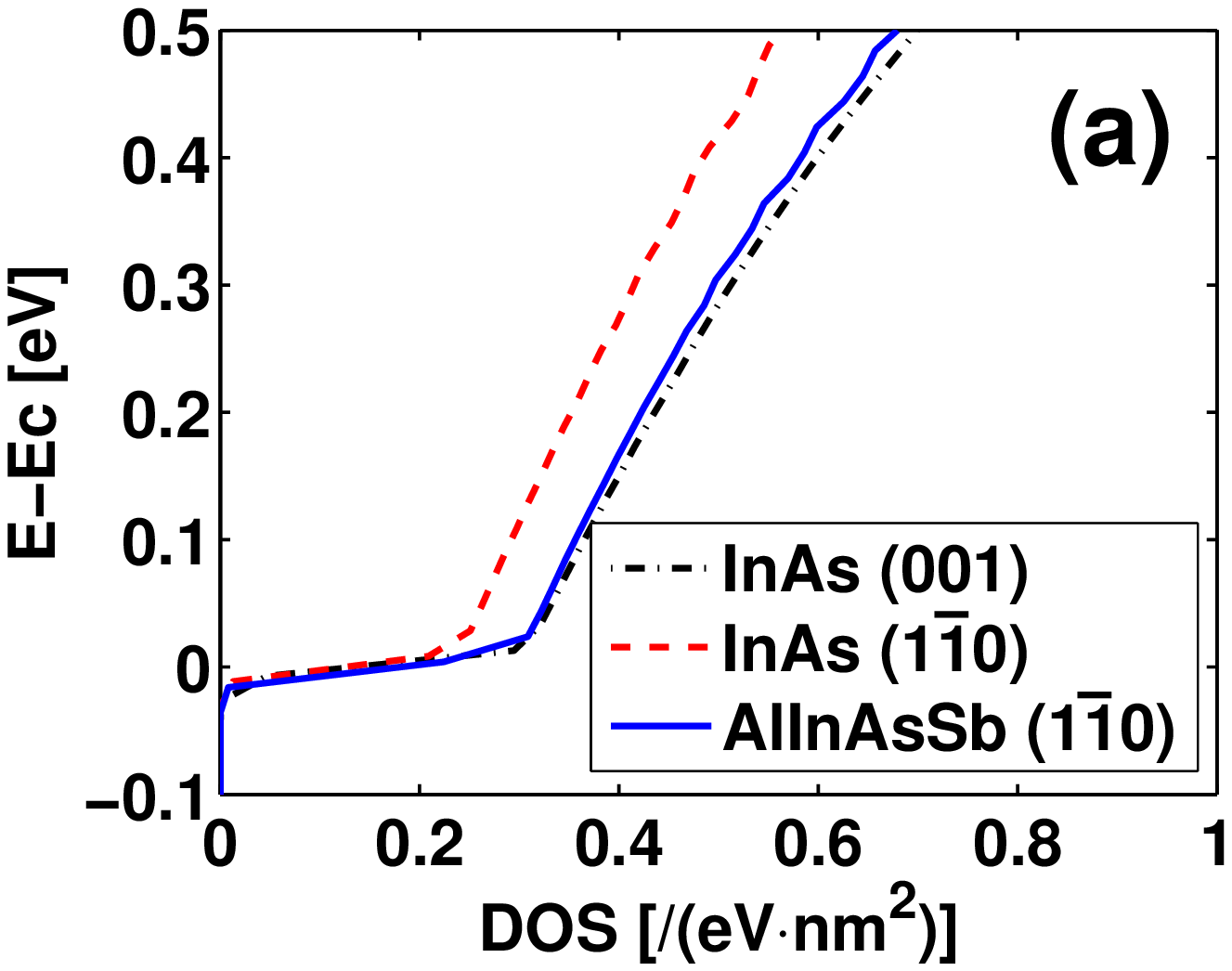}}
{\includegraphics[width=4.35cm]{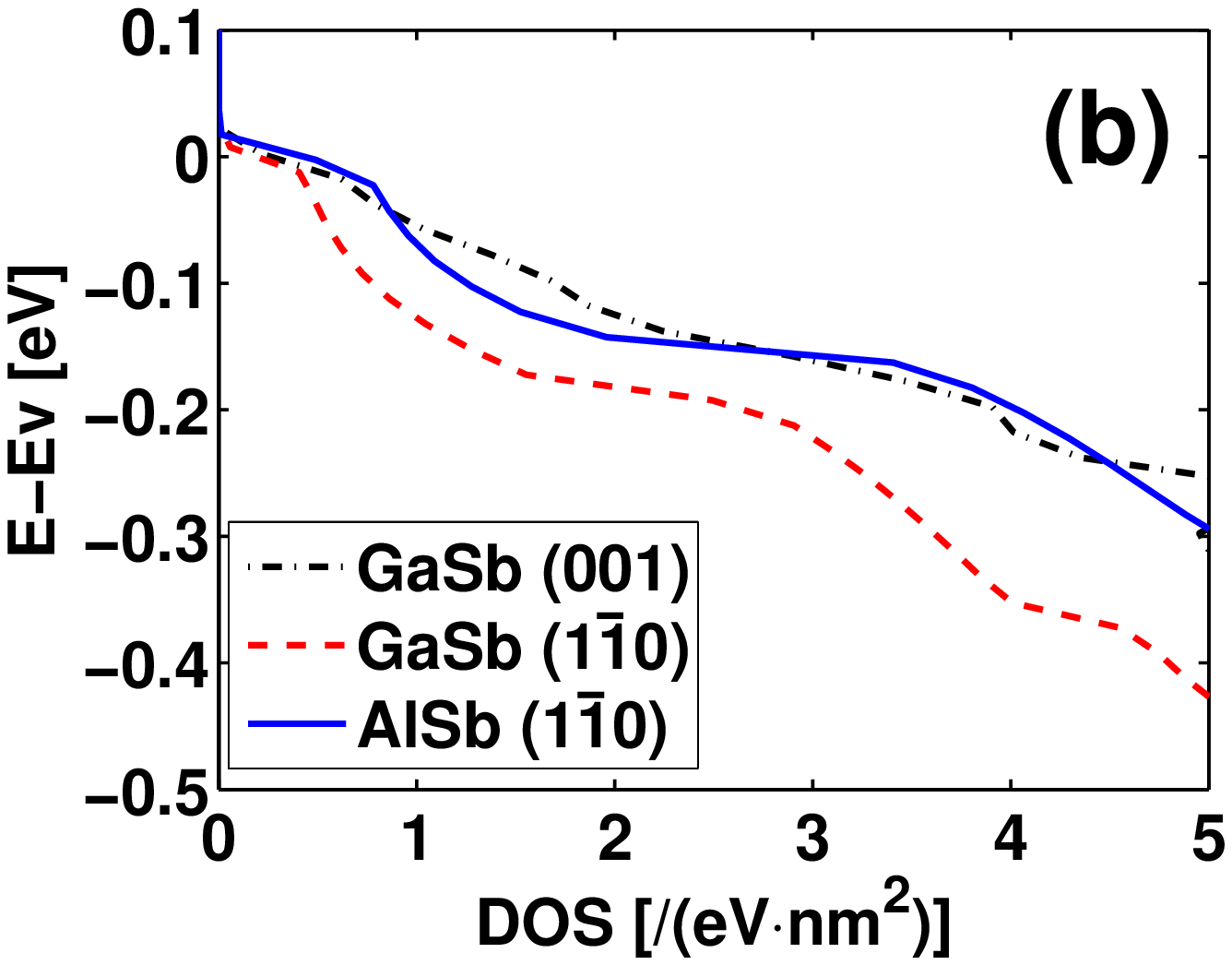}}
\caption{(a) Conduction and (b) valence band DOS of the six UTBs in Fig. \ref{fig:ek}.}
\label{fig:dos}
\end{figure}

The improvements can be understood from the band diagrams (Fig. \ref{fig:iv_band_trans} (c)) and transmission probabilities (Fig. \ref{fig:iv_band_trans} (d)). Compared with the (001)/[100] orientation, the (1$\bar{1}$0)/[110] orientation has larger transmission below the channel valence band edge (Ev), leading to larger $I_{\rm{ON}}$. However, its transmission above the channel Ev is also larger and the slope is less steep, leading to larger source-to-drain leakage and larger SS. As seen in the band structures plotted in Fig. \ref{fig:ek}, the (1$\bar{1}$0)/[110] InAs/GaSb UTB has smaller tunnel barrier energy and transport effective masses than the (001)/[100] InAs/GaSb UTB. Moreover, the source Fermi degeneracy, {\em i.e.}, the energy separation between the source Fermi level and the conduction band edge (Ec), is larger and the channel valence band DOS is smaller (Fig. \ref{fig:dos} (b)), changes which improve the superlinear onset and reduce the delayed saturation \cite{Taur2015,Wu2016,Rajamohanan2013}.

%\begin{table}
%\caption{Effective masses at conduction band and valence band edges.}
%\label{tab:masses}       % Give a unique label
%\begin{tabular}{p{1.7cm}p{1.3cm}p{2.2cm}p{2.1cm}}
%\hline\noalign{\smallskip}
%Material    &InAs (001) &InAs (-110) &AlSbInAs (-110) \\
%\hline\noalign{\smallskip}
%$m^*_e$ transport  & 0.1097  & 0.0514 (-53.1\%)   &  0.0698 (+35.8\%)  \\
%$m^*_e$ transverse & 0.1097  & 0.0706 (-35.6\%)   &  0.0789 (+11.8\%)  \\
%\hline\noalign{\smallskip}
%\hline\noalign{\smallskip}
%Material   &GaSb (001) &GaSb (-110) &AlSb (-110) \\
%\hline\noalign{\smallskip}
%$m^*_h$ transport  & -0.0658  & -0.0405 (-38.5\%)   &  -0.0791 (+95.3\%)  \\
%$m^*_h$ transverse & -0.0658  &-0.1635 (+148.5\%)  &  -0.2641 (+61.5\%) \\
%\hline\noalign{\smallskip}
%\end{tabular}
%\end{table}

\section{Triple-Heterojunction (3HJ) pTFET}
A 3HJ design is proposed to overcome the shortcomings of the (1$\bar{1}$0)/[110] HJ pTFET, i.e., the degraded SS and the small $I_{\rm{ON}}$.
The 3HJ pTFET here consists of an $\rm{(AlSb)}_{x2}\rm{(InAs)}_{1-x2}$ source, an $\rm{(AlSb)}_{x1}\rm{(InAs)}_{1-x1}$ source grade, an InAs source well, a GaSb channel well, an $\rm{Al}_{y1}\rm{Ga}_{1-y1}\rm{Sb}$ channel grade, and an $\rm{Al}_{y2}\rm{Ga}_{1-y2}\rm{Sb}$ channel/drain, all are lattice matched and aligned in the (1$\bar{1}$0)/[110] orientation (Fig. \ref{fig:device} (b)). The mole fractions x1, x2, y1, y2, and the region lengths L4 to L7 are the design parameters, which are optimized for the largest $I_{\rm{ON}}$ (Table \ref{tab:device_param}).

Fig. \ref{fig:iv_band_trans} (a) shows that the (1$\bar{1}$0)/[110] 3HJ design greatly improves the SS and $I_{\rm{ON}}$ of the (1$\bar{1}$0)/[110] HJ design, with $488\rm{A/m}$ ballistic $I_{\rm{ON}}$ obtained at $V_{\rm{DD}}=0.3\rm{V}$ and $I_{\rm{OFF}}=10^{-3}\rm{A/m}$. The reference InAs/GaSb HJ pTFETs show $1.4\rm{A/m}$ and $14.5\rm{A/m}$ ballistic $I_{\rm{ON}}$ respectively in the (001)/[100] and (1$\bar{1}$0)/[110] orientations. Fig. \ref{fig:iv_band_trans} (b) shows that the output $I_{DS}$-$V_{DS}$ characteristics are improved; comparing the (1$\bar{1}$0)/[110] 3HJ and (001)/[100] HJ designs, the onset (saturation) voltage is reduced from -0.070V (-0.267V) to -0.017V (-0.173V). Fig. \ref{fig:iv_band_trans} (c) and (d) show that the 3HJ design has a much thinner tunnel barrier and thus much larger tunneling probability (approaching unity) when turned on. Further, the 3HJ design shows a much steeper variation of transmission {\em vs.} energy above the channel Ev, implying less source-to-drain leakage and steeper turn-off characteristics.

From Fig. \ref{fig:ek} (c) and (e) it is observed that a (1$\bar{1}$0) AlInAsSb UTB has higher conduction band edge energy than a (1$\bar{1}$0) InAs UTB. This conduction band offset forms a quantum well in the source, which shortens the source depletion length and creates a resonant state above the well, both effects enhancing the tunneling probability. Further, the (1$\bar{1}$0) AlInAsSb UTB has larger electron effective masses (in both transport and transverse directions) than the (1$\bar{1}$0) InAs UTB, thus a larger conduction band DOS (Fig. \ref{fig:dos} (a)) and reduced source Fermi degeneracy (Fig. \ref{fig:iv_band_trans} (c)). From Fig. \ref{fig:ek} (d) and (f) it is found that the (1$\bar{1}$0) AlSb UTB has lower valence band edge than the (1$\bar{1}$0) GaSb UTB. This valence band offset forms a quantum well in the channel, which also shortens the tunnel barrier thickness and creates another resonant state below the well, both further enhancing the tunneling probability. Moreover, the AlSb UTB channel has larger hole effective masses than the GaSb UTB channel, leading to smaller source-to-drain leakage. Grading of the source HJ and channel HJ makes further improvements by further increasing the electric field at the tunnel junction and by tuning the positions of the resonant states. Note that, although the source Fermi degeneracy is reduced and the channel DOS is increased (Fig. \ref{fig:dos} (b)), the output characteristic is not degraded. This is due to the much higher transmission transparency enabled by the 3HJ design.

\begin{figure}[htbp] \centering
{\includegraphics[width=4.35cm]{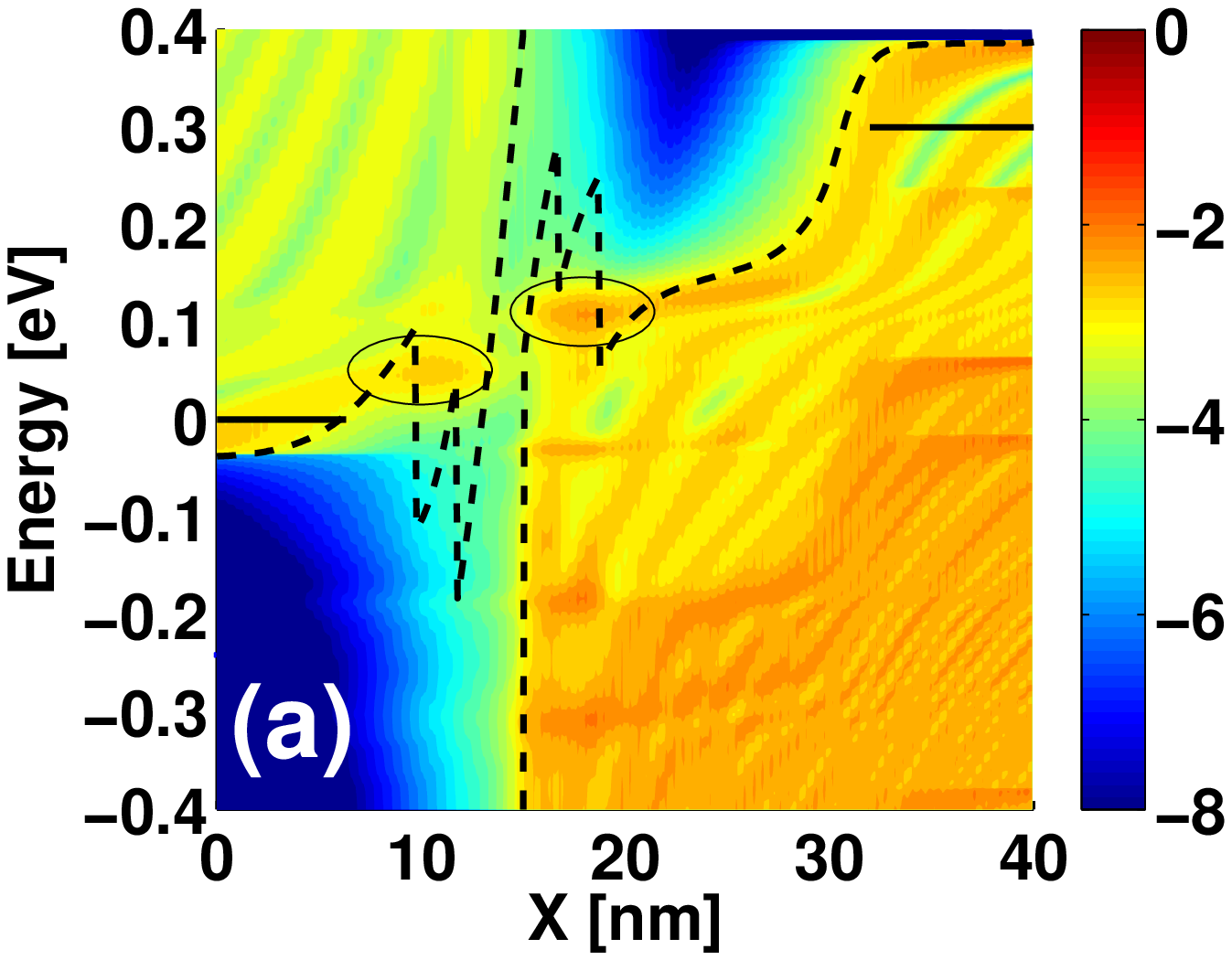}}
{\includegraphics[width=4.35cm]{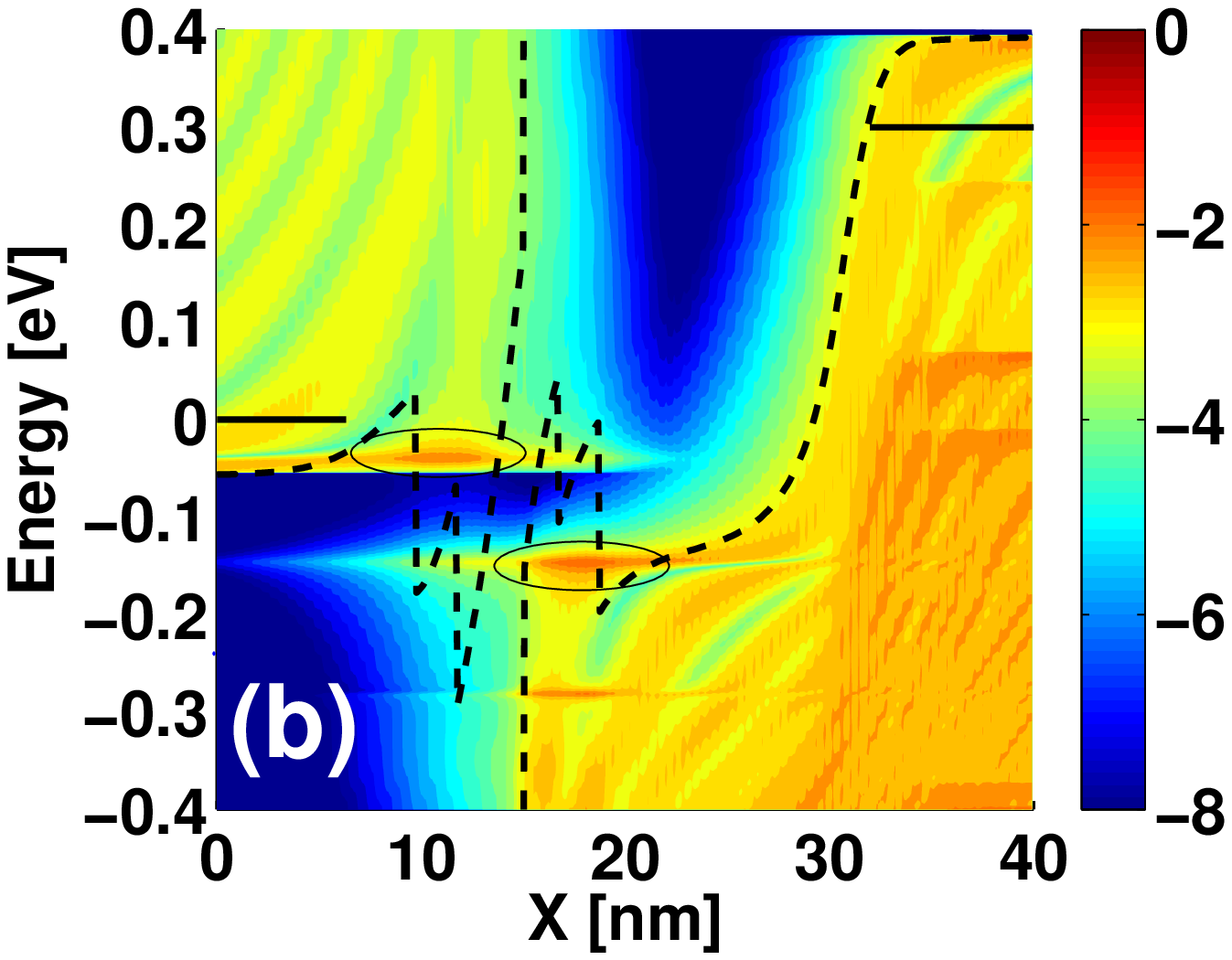}}
\caption{Logarithmic scale LDOS ($k_z=0$) of the 3HJ pTFET, at ON (a) and OFF (b) states. Band diagrams (dashed lines) and contact Fermi levels (solid lines) are superimposed. The quasi-bound states are highlighted with circles.}
\label{fig:ldos}
\end{figure}

Fig. \ref{fig:ldos} (a) and (b) depict the ON and OFF state local density of states (LDOS).
In the ON state, the two resonant states created by the two quantum wells both fall in the Fermi conduction window, enhancing the current. In the OFF state, there are no quasi-bound states inside the quantum wells, reducing the thermal emission induced leakage. However, because the tunnel barrier is so thin, evanescent states incident from the source (channel) could still couple to the propagating states of the channel (source) through interaction with phonons, forming a leakage current path that is not modeled here.

\begin{figure}[htbp] \centering
{\includegraphics[width=4.35cm]{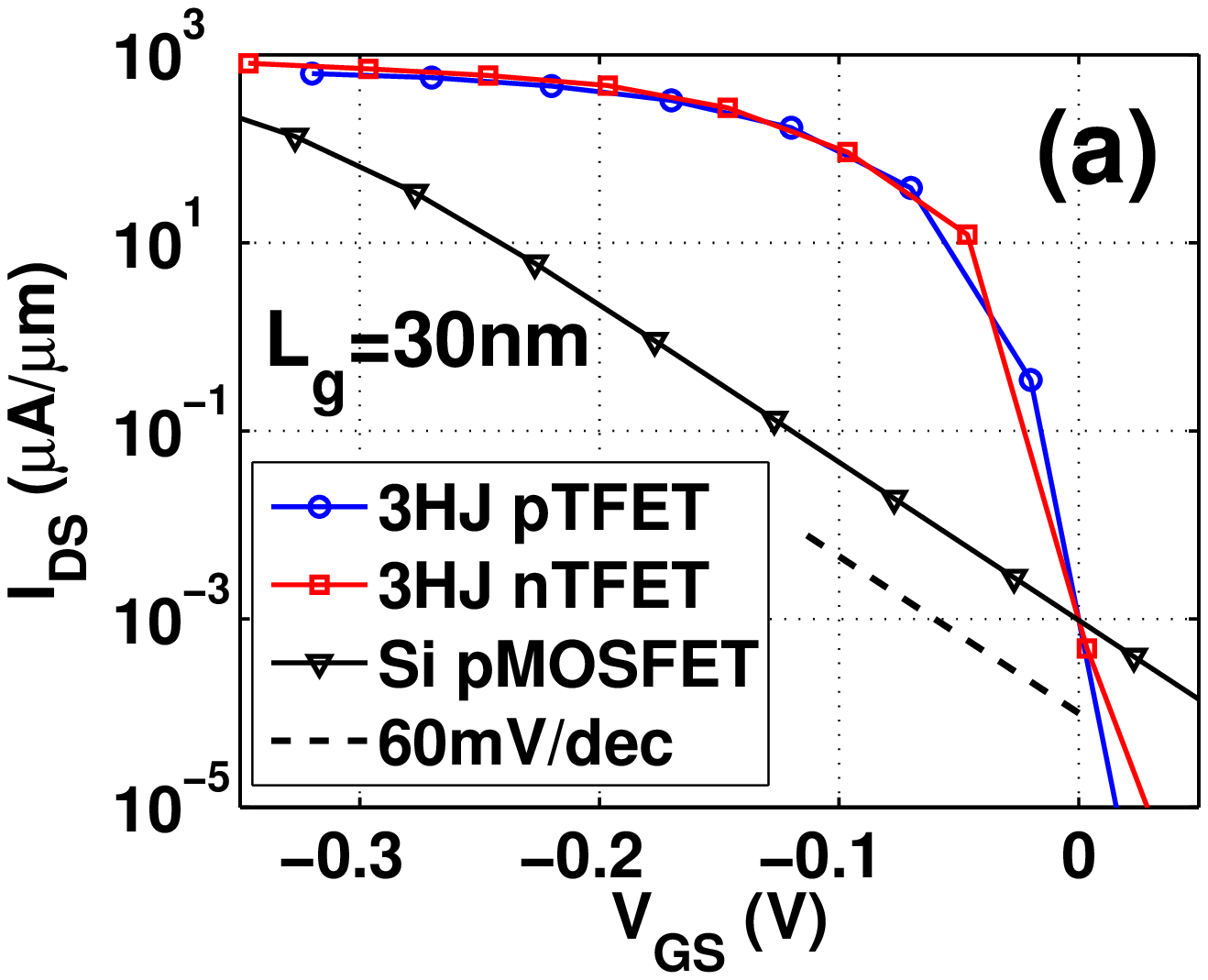}}
{\includegraphics[width=4.35cm]{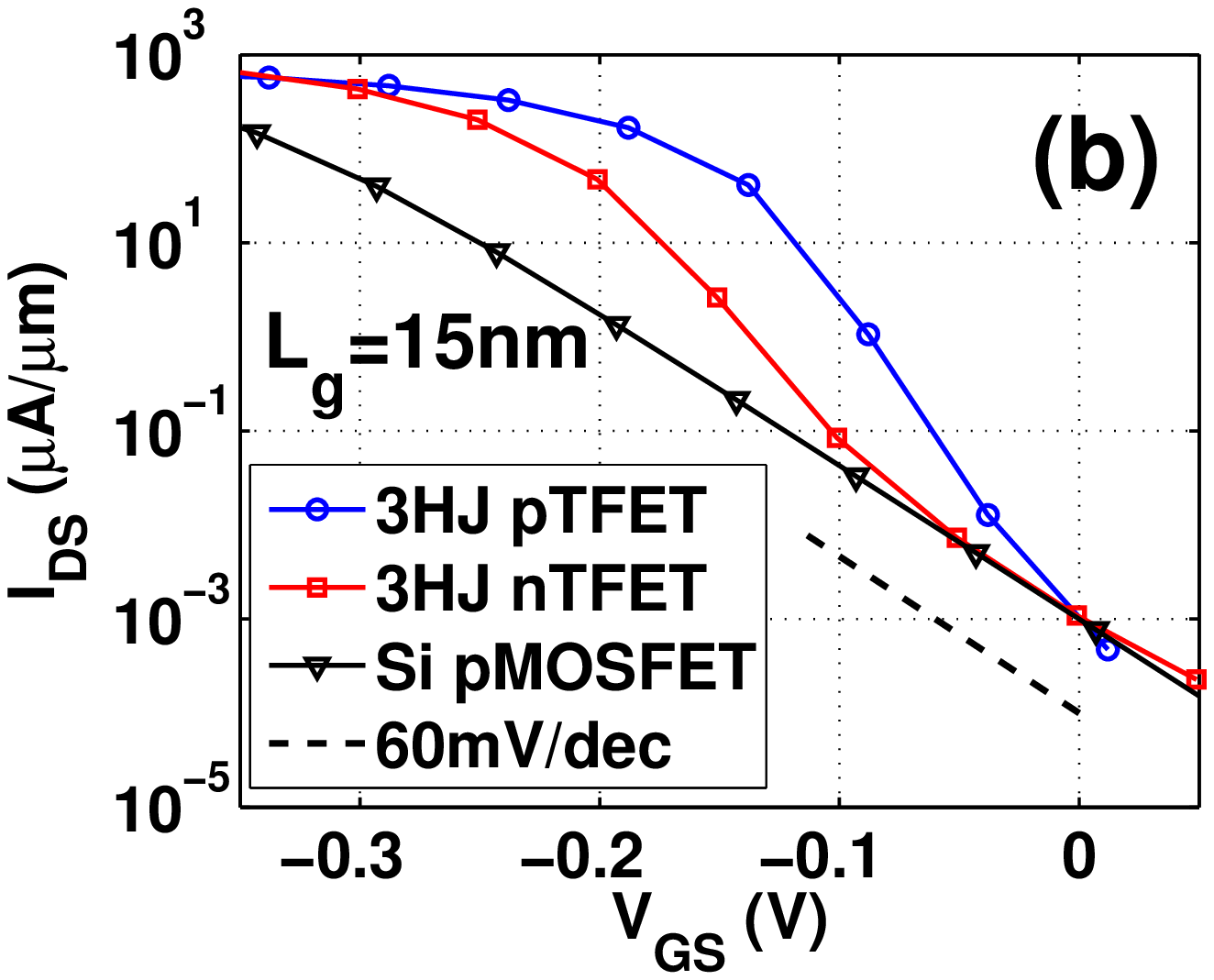}}
\caption{Transfer characteristics of the 3HJ pTFET, in comparison with the 3HJ nTFET and Si pMOSFET, at $L_g=30\rm{nm}$ (a) and $L_g=15\rm{nm}$ (b).}
\label{fig:compare_ntfet_mosfet}
\end{figure}

Finally, we compare the 3HJ pTFETs with corresponding 3HJ nTFETs (using the same materials and orientations) \cite{Long2016drc} and Si pMOSFETs (Fig. \ref{fig:compare_ntfet_mosfet}). For $30\rm{nm}$ ($15\rm{nm}$) channel length, $I_{\rm{ON}}$ of 3HJ pTFET is $582\rm{A/m}$ ($488\rm{A/m}$), comparable to the 3HJ nTFET and much larger than the Si pMOSFET. For 15nm channel length, the 3HJ pTFET has better SS and thus slightly larger $I_{\rm{ON}}$ than the 3HJ nTFET, owing to the larger channel band gap and channel effective mass of the 3HJ pTFET.

\section{Conclusion}
Design of III-V pTFETs is very challenging because of small source and large channel density of states.
By engineering crystal orientations and employing triple heterojunctions, very large ballistic ON currents are simulated for pTFETs, comparable to the n-type counterparts and significantly exceeding Si pMOSFETs. Improved linear and saturation regions are also observed in the output I-V characteristics. However, the large ballistic current may be degraded by phonon assisted tunneling, a topic of future study.

% or


%\appendices

%\section{}

%\section*{Acknowledgment}

\ifCLASSOPTIONcaptionsoff
  \newpage
\fi


\begin{thebibliography}{1}
\bibitem{ionescu2011tunnel}
A.~M. Ionescu and H.~Riel, ``Tunnel field-effect transistors as
energy-efficient electronic switches,'' \emph{Nature}, vol.~479, pp. 329--337, Nov. 2011. DOI: 10.1038/nature10679
\bibitem{mohata2011demonstration}
D. K. Mohata, R. Bijesh, S. Mujumdar, C. Eaton, R. Engel-Herbert, T. Mayer, V. Narayanan, J. M. Fastenau, D. Loubychev, A. K. Liu, and S. Datta ``Demonstration of MOSFET-like on-current performance in arsenide/antimonide tunnel FETs with staggered hetero-junctions for 300mV logic applications,'' \emph{IEEE IEDM Tech. Dig.}, Dec. 2011, pp.33.5.1--33.5.4. DOI: 10.1109/IEDM.2011.6131665
\bibitem{brocard2013design}
S.~Brocard, M.~G. Pala, and D.~Esseni, ``Design options for hetero-junction tunnel FETs with high on current and steep sub-threshold voltage slope," \emph{IEEE IEDM Tech. Dig.}, Dec. 2013, pp.5.4.1--5.4.4. DOI: 10.1109/IEDM.2013.6724567
\bibitem{Verreck2016}
D. Verreck, A. S. Verhulst, M. L. Van de Put, B. Sor\'{e}e, N. Collaert, A. Mocuta, A. Thean, and G. Groeseneken, ``Uniform strain in heterostructure tunnel field-effect transistors," \emph{IEEE Electron Device Lett.}, vol. 37, no. 3, pp. 337--340, Mar. 2016. DOI: 10.1109/LED.2016.2519681
\bibitem{avci2013heterojunction}
U. E. Avci and I. A. Young, ``Heterojunction TFET scaling and resonant-TFET for steep subthreshold slope at sub-9nm gate-length,'' \emph{IEEE IEDM Tech. Dig.}, Dec. 2013, pp.4.3.1--4.3.4. DOI: 10.1109/IEDM.2013.6724559
\bibitem{pala2015exploiting}
M. G. Pala, and S. Brocard, ``Exploiting hetero-junctions to improve the performance of III-V nanowire tunnel-FETs," \emph{IEEE J. Electron Devices Soc.}, vol.~3, no.~3, pp. 115--121, 2015. DOI: 10.1109/JEDS.2015.2395719
\bibitem{Long2016design}
P. Long, E. Wilson, J. Z. Huang, G. Klimeck, M. J.W. Rodwell, and M. Povolotskyi, ``Design and simulation of GaSb/InAs 2D transmission-enhanced tunneling FETs," \emph{IEEE Electron Device Lett.}, vol. 37, no. 1, pp. 107–-110, Jan. 2016. DOI: 10.1109/LED.2015.2497666
\bibitem{Ganapathi2011}
K. Ganapathi and S. Salahuddin, ``Heterojunction vertical band-to-band tunneling transistors for steep subthreshold swing and high on current," \emph{IEEE Electron Device Lett.}, vol. 32, no. 5, pp. 689--691, May 2011. DOI: 10.1109/LED.2011.2112753
\bibitem{brocard2014large}
S. Brocard, M. G. Pala, and D. Esseni, ``Large on-current enhancement in hetero-junction tunnel-FETs via molar fraction grading," \emph{IEEE Electron Devices Lett.}, vol.~35, no.~2, pp. 184--186, 2014. DOI: 10.1109/LED.2013.2295884
\bibitem{Li2015}
W. Li, S. Sharmin, H. Ilatikhameneh, R. Rahman, Y. Lu, J. Wang, X. Yan, A. Seabaugh, G. Klimeck, D. Jena, and P. Fay, ``Polarization-engineered III-nitride heterojunction tunnel field-effect transistors," \emph{IEEE J. Exploratory Solid-State Comput. Devices Circuits}, vol.~1, pp. 28--34, Dec. 2015. DOI: 10.1109/JXCDC.2015.2426433
\bibitem{Long2016}
P. Long, J. Z. Huang, M. Povolotskyi, G. Klimeck, and M. J.W. Rodwell, ``High-current tunneling FETs with (1$\bar{1}$0) orientation and a channel heterojunction," \emph{IEEE Electron Device Lett.}, vol. 37, no. 3, pp. 345–-348, Mar. 2016. DOI: 10.1109/LED.2016.2523269
\bibitem{Long2016drc}
P. Long, M. Povolotskyi, J. Z. Huang, H. Ilatikhameneh, T. Ameen, R. Rahman, T. Kubis, G. Klimeck, and M. J.W. Rodwell, ``Extremely high simulated ballistic currents in triple-heterojunction tunnel transistors," in \emph{Proc. 74th Annu. Device Res. Conf. (DRC)}, Jun. 2016.
\bibitem{Knoch2010}
J. Knoch and J. Appenzeller, ``Modeling of high-performance p-type III-V heterojunction tunnel FETs," \emph{IEEE Electron Device Lett.}, vol. 31, no. 4, pp. 305–-307, Apr. 2010. DOI: 10.1109/LED.2010.2041180
\bibitem{Avci2011}
U. E. Avci, R. Rios, K. J. Kuhn, and I. A. Young, ``Comparison of power and performance for the TFET and MOSFET and considerations for P-TFET," in \emph{Proc. 11th IEEE Int. Conf. Nanotechnol.}, Aug. 2011, pp. 869--872. DOI: 10.1109/NANO.2011.6144631
\bibitem{Avci2015}
U. E. Avci, D. H. Morris, and I. A. Young, ``Tunnel field-effect transistors: prospects and challenges," \emph{IEEE J. Electron Devices Soc.}, vol. 3, no. 3, pp. 88–-95, May 2015, DOI: 10.1109/JEDS.2015.2390591
\bibitem{Huang2015}
J. Z. Huang, L. Zhang, P. Long, M. Povolotskyi, and G. Klimeck, ``Quantum transport simulation of III-V TFETs with reduced-order
$\mathbf{k}\cdot\mathbf{p}$ method,"  in \emph{Tunneling Field Effect Transistor Technology}, L. Zhang and M. Chan, Eds. Switzerland: Springer, 2016, pp. 151--180. DOI: 10.1007/978-3-319-31653-6\_6
\bibitem{Verreck2014}
D. Verreck, A. S. Verhulst, B. Sor\'{e}e, N. Collaert, A. Mocuta, A. Thean, and G. Groeseneken, ``Improved source design for p-type tunnel field-effect transistors: Towards truly complementary logic," \emph{Appl. Phys. Lett.}, vol. 105, no. 5,
pp. 243506-1-–243506-4, Dec. 2014. DOI: 10.1063/1.4904712
\newpage
\bibitem{Taur2015}
Y. Taur, J. Wu, and J. Min, ``An analytic model for heterojunction tunnel FETs with exponential barrier," \emph{IEEE Trans. Electron Devices}, vol.~62, no.~5, pp. 1399--1404, 2015. DOI: 10.1109/TED.2015.2407695
\bibitem{Wu2016}
C. Wu, R. Huang, Q. Huang, J. Wang, and Y. Wang, ``Design guideline for complementary heterostructure tunnel FETs with steep slope and improved output behavior," \emph{IEEE Electron Device Lett.}, vol. 37, no. 1, pp. 20--23, Jan. 2016. DOI: 10.1109/LED.2015.2499183
\bibitem{Steiger2011}
S. Steiger, M. Povolotskyi, H.-H. Park, T. Kubis, and G. Klimeck, ``NEMO5: A parallel multiscale nanoelectronics
modeling tool," \emph{IEEE Trans. Nanotechnol.}, vol. 10, no. 6, pp. 1464–-1474, Nov. 2011. DOI: 10.1109/TNANO.2011.2166164
\bibitem{Luisier2006}
M. Luisier, A. Schenk, W. Fichtner, and G. Klimeck, ``Atomistic
simulation of nanowires in the $sp^3d^5s^*$ tight-binding formalism: From
boundary conditions to strain calculations," \emph{Phys. Rev. B}, vol. 74, no. 20, p. 205323. Nov. 2006. DOI: 10.1103/PhysRevB.74.205323
\bibitem{Tan2015}
Y. P. Tan, M. Povolotskyi, T. Kubis, T. B. Boykin, and G. Klimeck,
``Tight-binding analysis of Si and GaAs ultrathin bodies with subatomic
wave-function resolution," \emph{Phys. Rev. B}, vol. 92, no. 8, p. 085301, 2015.
DOI: 10.1103/PhysRevB.92.085301
\bibitem{Tan2016}
Y. Tan et al. \emph{Tight Binding Parameters by DFT Mapping.}
https://nanohub.org/resources/15173, accessed Apr. 26, 2016.
\bibitem{Rajamohanan2013}
B. Rajamohanan, D. Mohata, A. Ali, and S. Datta, ``Insight into the output characteristics of III-V tunneling field effect transistors," \emph{Appl. Phys. Lett.}, vol. 102, pp. 092105-1--092105-5, Mar. 2013. DOI: 10.1063/1.4794536
\end{thebibliography}
\end{document}